# Room-Temperature CsPbBr$_3$ Mixed Polaritons States

**VINCENT FORSTER,**[1,2,3,4] **SALVADOR ESCOBAR GUERRERO,**[5] **HUGO ALBERTO LARA-GARCÍA,**[5] **JEAN-FRANÇOIS BRYCHE,**[1,3,4] **ROCÍO NAVA,**[6] **DENIS MORRIS,**[1,2,3,4] **AND JORGE-ALEJANDRO REYES-ESQUEDA**[1,5,*]

[1]*Faculté des Sciences, Université de Sherbrooke, J1K 2R1 Sherbrooke, QC, Canada*
[2]*Institut Quantique, Université de Sherbrooke, J1K 2R1 Sherbrooke, QC, Canada*
[3]*Laboratoire Nanotechnologies Nanosystèmes (LN2), CNRS, Université de Sherbrooke, J1K 0A5 Sherbrooke, QC, Canada*
[4]*Institut Interdisciplinaire d'Innovation Technologique (3IT), Université de Sherbrooke, J1K 0A5 Sherbrooke, QC, Canada*
[5]*Instituto de Física, Universidad Nacional Autónoma de México, Circuito de la Investigación Científica, Ciudad Universitaria, Coyoacán, 04510, Ciudad de México, México*
[6]*Instituto de Energías Renovables, Universidad Nacional Autónoma de México, Privada Xochicalco s/n, Temixco, Morelos, 62580, México*
*\*reyes@fisica.unam.mx*

**Abstract:** Light-matter interactions are known to lead to the formation of polariton states through what is called strong coupling, leading to the formation of two hybrid states usually tagged as Upper and Lower Polaritons. Here, we consider a similar interaction between excitons and photons in the realm of strong interactions, with the difference that it enables us to obtain a mixed-polariton state. In this case, the energy of this mixed state is found between the energies of the exciton state and the cavity mode, resulting in an imaginary coupling coefficient related to a specific class of singular points. These mixed states are often considered unobservable, although they are predicted well when the dressed states of a two-level atom are considered. However, intense light confinement can be obtained by using a Bound State in the Continuum, reducing the damping rates, and enabling the observation of mixed states resulting from the correct kind of exceptional point giving place to strong coupling. In this study, using the Transfer Matrix Method, we simulated cavities made of porous silicon coupled with CsPbBr$_3$ perovskite quantum dots to numerically observe the mixed states as well as experimentally, by fabricating appropriate samples. The dispersion relation of the mixed states is fitted using the same equation as that used for strong coupling but considering a complex coupling coefficient, which is directly related to the appropriate type of exceptional point.

## 1. Introduction

First proposed around 1960 [1], strong coupling and the resulting bosonic quasiparticles known as polaritons have seen ever-increasing interest in the last few years, as several groups have raised their capacity to obtain and manipulate them by strongly controlling the parameters of the materials used to observe this coupling [2]. Following this achievement, new devices based on polariton Bose-Einstein condensation [3], polariton lasing and sensing [4,5], and modulating chemical reactions [6] have emerged, as well as all the possibilities for fabricating robust quantum technologies, given the drastic reduction in the future polariton-chip size and increase in their speed performance [7,8].

On the other hand, the ability to fabricate materials with specific parameters has also allowed observation in the optics of Bound States in the Continuum (BICs) for no more than a decade [9]. BICs allow strong localization of energy in an open resonator coupled to a radiation continuum [9-11], possessing infinite radiative lifetimes, thus enabling boundless enhancement of electric and magnetic fields through a divergent quality factor [12]. However, in practice, finite material extension, intrinsic absorption losses, fabrication defects, and structural disorders result in the fabrication of real-world leaky structures with quasi-BICs [13]. Two main types of quasi-BICs have attracted attention in recent years: symmetry-protected BICs and accidental BICs [14-16]. Given that a BIC can be explained by destructive interference, with only a discretized diffraction channel remaining open when photonic structures are on the subwavelength scale, the former is observed when the coupling of this channel to free space ceases owing to symmetry mismatch, whereas the latter can be observed by continuous parameter tuning [12,17].

The presence of BICs goes hand-in-hand with non-Hermitian physics because an open resonator implies a complex system with loss. In principle, a non-Hermitian system shows real eigenvalues if its Hamiltonian satisfies the conditions for parity-time (PT) symmetry [18]. However, in these systems, there are regions where symmetry is preserved and broken, resulting in the existence of Exceptional Points (EPs) and the degeneration of the eigenfunctions of the Hamiltonian and complex eigenvalues [19]. This eigenvalue coalescence implies that the Hamiltonian can no longer be diagonalized. In these cases, the system can be physically manipulated to be either at the EP singularity or encircle it [20]. In this direction, optics and photonics have become natural realms for experimentally observing and using these singularities [21,22].

BICs are often associated with emissive systems that improve or modify their emission properties. In this regard, perovskite nanocrystals have recently attracted attention owing to their potential applications in photovoltaic cells, light-emitting diodes, and tunable single-photon sources [23-27]. Similarly, the search for strong excitonic effects, low-temperature processing, and simple scalability, while allowing strong light-matter interactions, has also attracted significant interest in these materials to study room-temperature exciton-polariton formation [28-31]. In parallel, there is a growing interest in using porous silicon (p-Si) in biosensing and photonic applications, including strong coupling [32-39]. Indeed, porous silicon is a versatile nanoplatform thanks to the coral-like morphology produced by the electrochemical etching of crystalline silicon (c-Si). When extremely porous, light can be emitted because of the quantum confinement of charge carriers in the nanostructure as the p-Si skeleton thickness is reduced to a few nanometers (<4.3 nm, exciton Bohr radius) with the participation of surface states [40]. In addition, given that the etching process is self-limiting and primarily occurs at the pore tips, it opens the possibility of fabricating multilayered porous structures with a high contrast index between the porous layers, interface quality, and a large surface area (540-840 $m^2/cm^3$).

In this work, we discuss the theoretical frameworks used first to calculate the photonic response of the coupled system, corresponding to the p-Si photonic cavity and the $CsPbBr_3$ perovskite quantum dots (QDs), and then the model to calculate the coupling between them, that is, the fitting of the corresponding dispersion relation. In this frame, we discuss how the presence of BIC modes allows the manifestation of related Exceptional Points, which gives rise to strong coupling such that mixed polaritonic states can be observed. We then present our simulation results for a wide range of related parameters, based on the corresponding empirical parameterizing equation. From there, it follows naturally the fitting of the corresponding dispersion relations by considering the singularities resulting from the presence of BICs and EPs, that is, an imaginary coupling coefficient. We then discuss the band structure of the coupled system and topological charge of the mixed polaritonic states. Finally, we experimentally verified the results for some associated parameters.

## 2. Methods and protocols

*2.1 Transfer Matrix Method*

To study light propagation in stratified media, once Maxwell's equations and the appropriate boundary conditions have been established, the best approach is to calculate the Fresnel coefficients, and use the complete refractive index of the system to consider absorption and its possible effects on polarization. An effective and well-known way to achieve this is to use the Transfer Matrix Method (TMM) [41]. This methodology also allows the calculation of the reflectance, density of states, and field intensity of different layers of the media. By correctly introducing a gain, possible emissions from the media can also be considered. The power of this methodology has helped to study in detail the effects of asymmetry in the field localization and transmittance for supported microcavities [37] and has already been used in polariton formation and interaction in p-Si at room-temperature [38]. In this study, the TMM was used to simulate the photonic response of a coupled system consisting of an asymmetric cavity and a layer of perovskite quantum dots.

## 2.2 Strong coupling model

The interaction of light with matter can be described under some approximations by a Jaynes-Cummings Hamiltonian as $H_0 = \sum_k E_X(k) b_k^\dagger b_k + \sum_k E_C(k) a_k^\dagger a_k + \sum_k g\, (a_k^\dagger b_k + b_k^\dagger a_k)$, where $E_X$ is the exciton energy, $E_C$ is the photon energy in the cavity, and $g$ is the coupling term, which is better known as Rabi splitting. The first term represents matter corresponding to the electron states of a generic two-level system, the second term corresponds to a structured radiation field, and the third term describes the interaction between the two systems [42]. In a weak-coupling situation, the interaction can be described by the photon excitation of an electron from the ground state to the excited state. When coupling becomes more important, the structured light field and specific optical transitions are adequately described by the corresponding creation and annihilation operators, as stated by the Hamiltonian above. Within this context, it is normal to search for many quantum emitters or a small mode volume-confining light to achieve large coupling strengths [43].

However, by using the simpler and more intuitive non-Hermitian Hamiltonian for coupled oscillators, it is also possible to analyze in a richer way the presence of BICs and EPs in a given system because this Hamiltonian is not diagonalizable at the critical coupling strength, given by $g_{QEP} = |\gamma_{cav} - \gamma_{ex}|/4$, where the quantum exceptional point of this coupled system gives place to Rabi splitting. In addition, as established in [44], under correct assumptions, the approach to this Hamiltonian indicates that the strong coupling phenomenon can be treated classically as well as quantum-mechanically, using the following equations:

The Hamiltonian can be written as:

$$H = \begin{pmatrix} \omega_{cav} - i\frac{\gamma_{cav}}{2} & g \\ g & \omega_{ex} - i\frac{\gamma_{ex}}{2} \end{pmatrix}, \tag{1}$$

with the corresponding eigenvalue problem and polaritonic eigenvalues:

$$\begin{pmatrix} \omega_{cav} - i\frac{\gamma_{cav}}{2} & g \\ g & \omega_{ex} - i\frac{\gamma_{ex}}{2} \end{pmatrix} \begin{pmatrix} C \\ X \end{pmatrix} = \omega_\pm \begin{pmatrix} C \\ X \end{pmatrix}, \tag{2}$$

$$\omega_\pm = \frac{\omega_{cav} + \omega_{ex}}{2} - \frac{i}{2}\left(\frac{\gamma_{cav}}{2} + \frac{\gamma_{ex}}{2}\right) \pm \sqrt{g^2 + \frac{1}{4}\left(\delta - i\left(\frac{\gamma_{cav}}{2} - \frac{\gamma_{ex}}{2}\right)\right)^2}, \tag{3}$$

where $\omega_{cav}$ and $\omega_{ex}$ are the uncoupled cavity mode and QDs exciton angular frequencies, respectively; $\gamma_{cav}$ and $\gamma_{ex}$ are the damping rates of the two states; $\omega_+$ and $\omega_-$ are the frequencies of the hybrid states, which usually are recognized as the high- and low-polariton branches, respectively. $g$ is the coupling strength, and $\delta = \omega_{cav}(\theta) - \omega_{ex}$ is the angular or in-plane $k$-detuning between the cavity and QDs exciton energies, considering the angle-resolved cavity mode. $C$ and $X$ are the Hopfield coefficients representing the weighting coefficients of the cavity mode and exciton for each hybrid state, respectively, where $|C|^2 +$

$|X|^2 = 1$. The energy separation between the hybrid polariton bands at the anti-crossing, $\delta = 0$, defines mode splitting or Rabi splitting as follows:

$$\Omega = \sqrt{4g^2 - \left(\frac{\gamma_{cav}}{2} - \frac{\gamma_{ex}}{2}\right)^2}. \qquad (4)$$

Apart from $\delta$ or $k$-detuning, strong coupling is also regularly studied in terms of the energy difference among the cavity and the exciton at normal incidence, that is, the cavity-detuning $\Delta = (\omega_{cav} - \omega_{ex})_{\theta=0}$, where three cases are possible: negative $(\omega_{cav} - \omega_{ex})_{\theta=0} < 0$, positive $(\omega_{cav} - \omega_{ex})_{\theta=0} > 0$, or zero $(\omega_{cav} - \omega_{ex})_{\theta=0} = 0$ detuning.

To discuss the obtained results in terms of BICs and EPs, two considerations must be made. The first concerns the conditions required for non-vanishing Rabi splitting and spectrally separable resonances, given by $2|g| > \left|\frac{\gamma_{cav}}{2} - \frac{\gamma_{ex}}{2}\right|$ and $\Omega > \frac{\gamma_{cav}}{2} + \frac{\gamma_{ex}}{2}$, respectively [45]. The second is the condition for the occurrence of an EP, that is, when the square-root term in Eq. (3) is zero because the two eigenvalues coalesce [21]. As mentioned above, Rabi splitting occurs from this EP when a real coupling constant is assumed. However, as previously discussed [44], and we will show below, for certain cases, more free considerations allowed by the non-Hermitian character of the system grant the coupling constant to be imaginary, and then the presence of mixed polaritonic states.

*2.3 Perovskites QDs fabrication*

Regarding the perovskite QDs preparation, by following a typical hot-injection synthesis, 0.2 mmol of PbBr2 (73.4 mg) and 5 ml of 1-octadecene were loaded in a 25 ml three-neck round bottom flask, the mixture was nitrogen purged at 120 °C for 1 hour, then 0.5 ml of oleylamine and 0.5 ml of oleic acid were injected into the reaction flask, the mixture was stirred until all PbBr2 was completely dissolved. Then 0.2 ml of OLA-HBr at 80 °C and 0.5 ml of Cs-oleate at 100 °C were injected successively. The reaction mixture was cooled in an ice bath immediately after the final injection. Then, 5 ml of the crude solution was loaded into a centrifuge tube containing 5 ml of acetone and centrifugated at 7000 rpm for 30 min. After centrifugation, the supernatant was discarded, and the precipitate was dispersed in toluene. In this way, QDs approximately 8-10 nm in size can be obtained [46].

*2.4 p-Si cavities fabrication, complex refractive index calculation and photoluminescence measurement setup*

Porous silicon (p-Si) microcavities were fabricated by electrochemical etching of p-type boron-doped crystalline silicon (c-Si) wafers with a (100) orientation and an electrical resistivity < 0.005 Ω cm. Before the etching process, an aluminum layer was evaporated on the backside of the c-Si wafers and heated to 550 ºC in an inert atmosphere for 15 min, to create an electrical contact. The Teflon cell was filled with an electrolyte composed of aqueous hydrofluoric acid (HF), ethanol, and glycerin at a volume ratio of 3:7:1. Electrochemical etching, in which a c-Si substrate was the cathode and a platinum mesh was used as the anode, was initiated by applying a constant electrical current. The porosity and thickness of the p-Si layer depend on the current density and etching time, respectively, both of which are controlled by a computer and Keithley 2450 Source Meter SMU Series. This control of the porosity and thickness determines the optical path length of each layer, allowing the tuning of its photonic response. To minimize the porosity gradient in each layer, pauses of 1 s every 4 s of etching were applied during the anodization. p-Si simple bilayers with low and high porosities were then fabricated by alternating the applied current density during the electrochemical etching, between two values, 3 mA/cm$^2$ and 40 mA/cm$^2$, respectively, denoted as layers A and B. Multiple layer stack samples, or microcavities, were obtained by combining low- and high-porosity layers in a Distributed Bragg Reflector (DBR) sequence, where the conditions for the respective thicknesses, $d_A$ and $d_B$, define their corresponding refraction indexes: $d_A = \frac{\lambda_{DBR}}{4n_A}, d_B = \frac{\lambda_{DBR}}{4n_B}$. After anodization, the samples were rinsed with ethanol and dried under a nitrogen flow. Finally, the p-Si samples were passivated by thermal oxidation at 300 °C for 30 min. Previously, a layer of aluminum was removed from the c-Si substrate to prevent its diffusion during thermal oxidation and to avoid possible Al contamination during the quantum dots deposition.

P-Si is a nano-structured material composed of a skeleton of c-Si surrounded by air. The complex refractive index, $\eta = n - ik$, can be calculated using the effective medium approximation (EMA), following Estrada-Wise and del Río [47], as shown in [38].

Angle-resolved photoluminescence (ARPL) was measured at the University Laboratory of Optics at Surfaces (Laboratorio Universitario de Óptica de Superficies) into the Physics Institute of UNAM (LOS-UNAM). The sample was excited at 355 nm using an EKSPLA PL2231-50-SH/TH Nd:YAG pulsed laser System featuring ~26 ps pulses with a repetition rate of 10 Hz. The spot diameter was fixed at 3.6 mm by using a diaphragm. The emission of the sample was collected using an optical fiber (Ocean Optics model P1000-2-UV-VIS with a core of 1000 µm) and analyzed using an Ocean Optics USB2000+ UV-VIS spectrometer. Both the sample and optical fiber were attached to a rotational Newport RSP-1T plate to control the incidence and emission angles, respectively.

## 3. Results and discussion

### 3.1 qBICs and EPs

Recently, we discussed the presence of qBICs in asymmetric photonic crystals [37] and aperiodic quasi-crystals [38]. When considering intrinsic-lossy, non-Hermitian systems, such as p-Si, the introduction of an asymmetry allows an increase in the transmission and field localization of the formerly symmetric micro-cavity [37]. This result is directly related to the appearance of robust states, with practically zero bandwidth and large factor quality, giving rise to a broadband large Purcell factor, that is, the previously discussed qBICs [48-50]. In this study, we selected an asymmetric mirror cavity as the photonic component of the coupled system. This cavity, as shown in Fig. 1(a), consists of ten pairs of alternating layers with different refractive properties, A and B, with a defined thickness of each one, an extra layer of type A at the bottom to break the mirror cavity, and a layer called a defect of type B, whose thickness can be varied at the top. Fig. 1(a) shows the selected design, while Fig. 1(b) shows the reflectance at normal incidence ($k = 0$), and the corresponding density of states, which can be obtained for a given set of fabrication parameters, considering a non-zero cavity-detuning. The importance of the fabrication parameters is discussed in detail in the next section when discussing mixed polaritons.

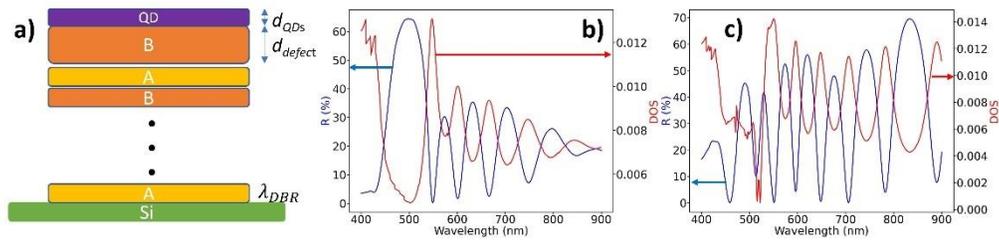

Fig. 1. (a) Asymmetric mirror cavity design, normal incidence reflectance (b) with and (c) without the perovskite quantum dots layer.

The large density of states and the corresponding field localization are evidence of the BIC nature of the cavity modes that can be obtained for this design. Further proof can be obtained when looking at the reflectance at normal incidence ($k = 0$) from the coupled system, as shown in Fig. 1(c), where thinner and deeper modes are formed because of the strong coupling between the perovskite quantum dots, as shown in Fig. 1(a), and photonic array. This coupled system was obtained by depositing a layer of perovskite QDs on top of the asymmetric mirror cavity, that is, on the top defect layer B. The thickness of the QDs layer was also modified. These modes from the coupled system would have a typical basic $Q$ factor of $Q = \lambda_{pol-BIC}/\gamma_{pol-BIC} = 2.2\times10^6$, where $\lambda_{pol-BIC} = 523.9$ nm and $\gamma_{pol-BIC} = 0.24$ pm, which is obtained from the fitting of this mode to a Lorentzian curve. As shown in the next section, the presence of qBICs in the reflectance of the coupled system translates to a divergence of its emission at the corresponding $k$ positions. This is a result of the large Purcell factor and the corresponding smaller decay rate associated with the BIC. However, there is another immediate consequence: the damping of this mode and the photoluminescence (PL) peak decrease to almost zero, becoming almost equal to each other.

If $\gamma_{cav} = \gamma_{ex}$, and given that the emission intensity is divergent or very high, which can be also understood as a population inversion condition, for a system where the cavity-detuning is non zero, then the EPs giving rise to a coupling condition are those identified as EP2 in Ref. [22] and the corresponding coupling constant becomes imaginary. Consequently, as discussed in the next section, the complex coupling constant is the only possible fit for the dispersion relations of the polaritons obtained, resulting in mixed polaritons.

It is necessary to say that mixed polaritons, surging from this kind of EPs, had already been predicted before under different considerations [51], which are satisfied here, given that the non-Hermitian Hamiltonian from Eq. (1) can be classified as a Hamiltonian of the von Neumann-Wigner type, characterized by a bounded potential [52]. From these previous studies, it can be concluded that the resulting mixed polaritons are also BICs, that is, pol-BICs, which are the corresponding eigenfunctions of this type of EPs.

### 3.2 Mixed polaritons

Based on previous results [37,38], to understand the influence of qBICs from the cavity modes on the formation of polaritons, we decided to study a simpler design, as presented in Section 3.1, and to study pol-BIC formation when varying the defect and QDs layer thicknesses $d_{defect}$ and $d_{QDs}$, as well as the value of $\lambda_{DBR}$, defined by the condition for the DBR sequence $n_A d_A = n_B d_B = \lambda_{DBR}/4$, for values not necessarily close to the exciton wavelength emission. The consequences of these variations on the light-matter interaction for this coupled system were studied by implementing Python code based on the TMM, as mentioned previously.

Before analyzing the simulation results, it is worth discussing the angle-resolved reflectance spectra in momentum space, that is, the reflectance as a function of the wavevector $k$. This is done for the asymmetric mirror cavity shown in Fig. 1, before, Fig. 2(a), and after, Fig. 2(b), the strong coupling with the perovskite QDs for specific values of $d_{defect}$ and $\lambda_{DBR}$, which is clearly different from exciton wavelength emission. As shown in Fig. 2(a), for values of $k$ of approximately $\pm 1\times10^7$ nm$^{-1}$, (angular range $\pm40$-$60$ °), a qBIC mode is clearly observed. In contrast, after the hybridization of the modes owing to strong coupling with the perovskite QDs, Fig. 2(b) shows the formation of pol-BICs linked to the previous qBIC. These pol-BICs have different characteristics depending on their energy with respect to the exciton energy and can be classified in the traditional way as Lower- (LP) or Upper-polaritons (UP). For instance, LPs are more intense and exhibit a narrower energy dispersion.

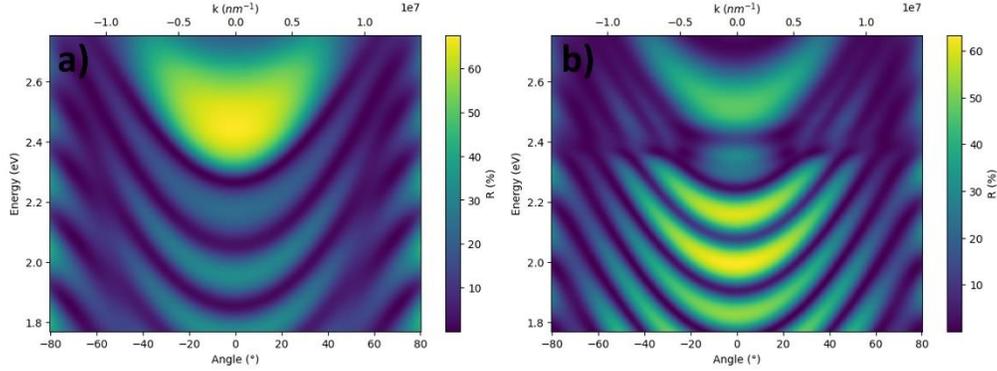

Fig. 2. Simulation of the Angle-resolved reflectance for the asymmetric mirror cavity (a) before, and (b) after the strong coupling with the perovskite quantum dots layer.

Fig. 3(a) shows the experimental photoluminescence (PL) of perovskite QDs in solution and deposited on p-Si, while Fig. 3(b) shows the simulated characteristic emission from a pol-BIC from this coupled system, which would be very intense, with a very small $\gamma_{pol-BIC}$. Therefore, in this study, the strong coupling and pol-BIC formation in this coupled system were followed by analyzing the PL instead of its reflectance.

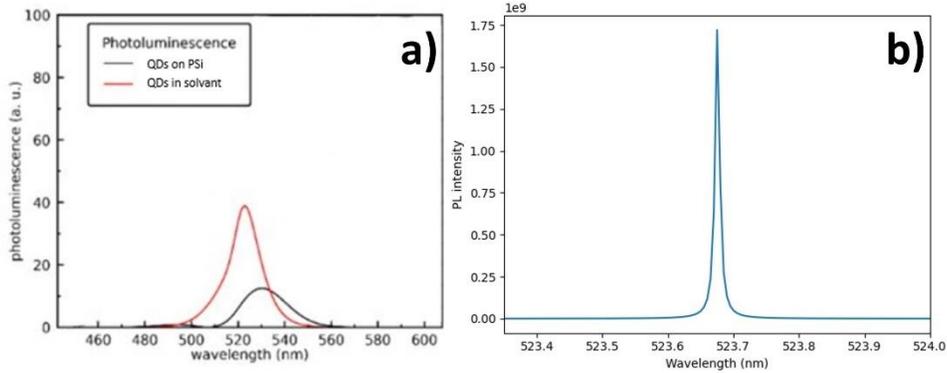

Fig. 3. Experimental emission from the perovskite QDs in (a) solution, on a p-Si layer, and (b) simulated emission when forming a pol-BIC.

Fig. 4 presents a cartography of the maximal intensity of the PL from the coupled system, for a range of values for $d_{defect}$ and $d_{QDs}$, respectively, regardless of the wavevector value, in two wavelength regions, [500, 525] and [525, 540], corresponding to UP and LP, respectively, where 525 nm is the wavelength emission of the non-coupled CsPbBr$_3$ perovskite system.

Figs. 4(a) and 4(c) show the UP emission with and without considering absorption from the p-Si and perovskite systems, respectively, whereas Figs. 4(b) and 4(d) show the LP emission with and without considering absorption from the p-Si and perovskite systems, respectively.

Surprisingly, the emission from the UP region became more intense when the absorption was considered. Although counter-intuitive, this is a manifestation of the qBIC mode, because it facilitates destructive interference between different modes with radiative dissipation. This effect is less evident for the LP part because these polaritons are already intense when the absorption is not considered (Fig. 2(b)). However, considering the patterns formed for both UP and LP emissions, there seems to be a shift when considering the absorption. To analyze this, attention should be focused on the region of 0-100 nm for $d_{QDs}$, varying $d_{defect}$, for the

UP, with and without absorption, that is, the left side of Figs. 4(a) and 4(c), respectively. From these figures, a periodicity of approximately 215 nm was established. By considering the refractive index of the defect layer, this periodicity becomes $\delta = n_B * L \approx 1.22 * 215 \approx 260.2\ nm \approx \frac{525}{2} \rightarrow \frac{\lambda_{exciton}}{2}$.

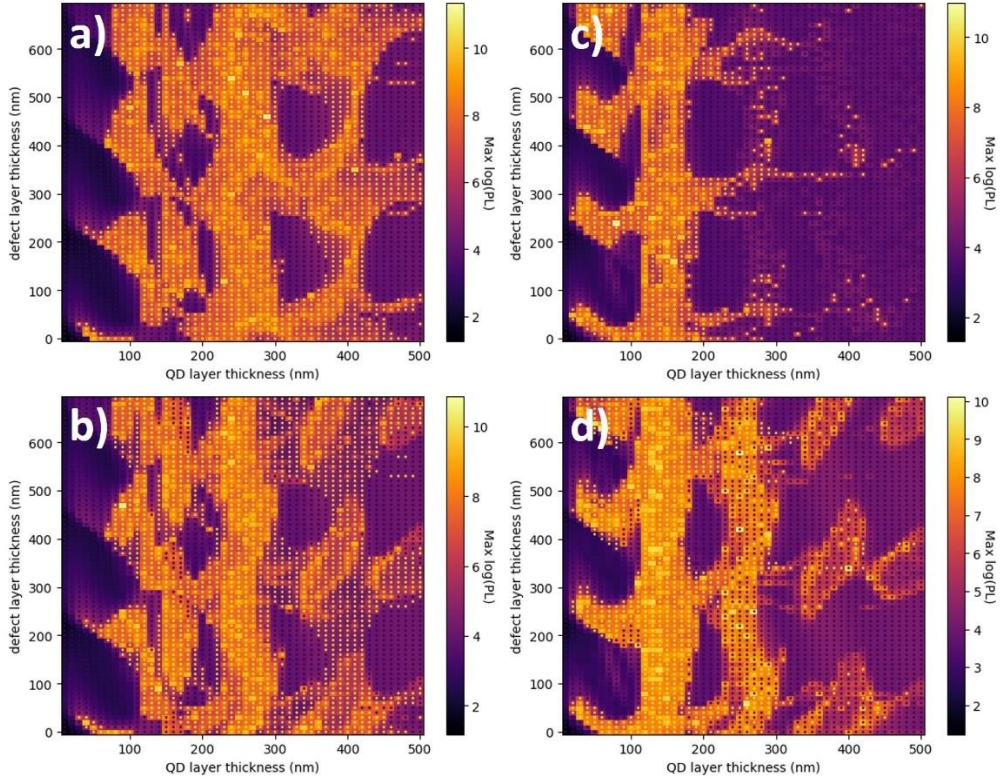

Fig. 4. Cartography of the maximal intensity of the simulated PL from the coupled system when varying $d_{defect}$ and $d_{QDs}$. (a) UP with absorption, (b) LP with absorption, (c) UP without absorption, and (d) LP without absorption.

To gain more insight into this result and the effect of qBICs presence in pol-BIC formation, that is, to determine the relation of dispersion for pol-BICs in this coupled system, one possible path is to fix $d_{defect}$, and to follow $k$ (angular position) for the possible pol-BICs when systematically varying $\lambda_{DBR}$ and $d_{QDs}$. The criterion to determine pol-BIC formation was to obtain a concentrated polariton emission, that is, an emission with an intensity above $10^5$. Fig. 5 shows the simulation results on a logarithmic scale. First, for a given $\lambda_{DBR}$, in analyzing the corresponding emission spectra, it could be observed that UPs were formed for minimal reflectance values, while LPs showed the opposite, that is, maximal reflectance values, as shown in Figs. 5(a) and 5(b). This directly corresponds to the results shown in Fig. 4. Because absorption minimizes reflectance, UPs are more intense when absorption is considered (Fig. 4(a)), whereas LPs are more intense when there is no absorption, that is, when the reflectance is higher, as shown in Fig. 4(b). A second observation obtained from this analysis is the coexistence of several UPs and/or LPs that are likely to be separated by the reflectance mode, as shown in Fig. 5(c), because several exist for certain cavity parameters, as shown in Fig. 1.

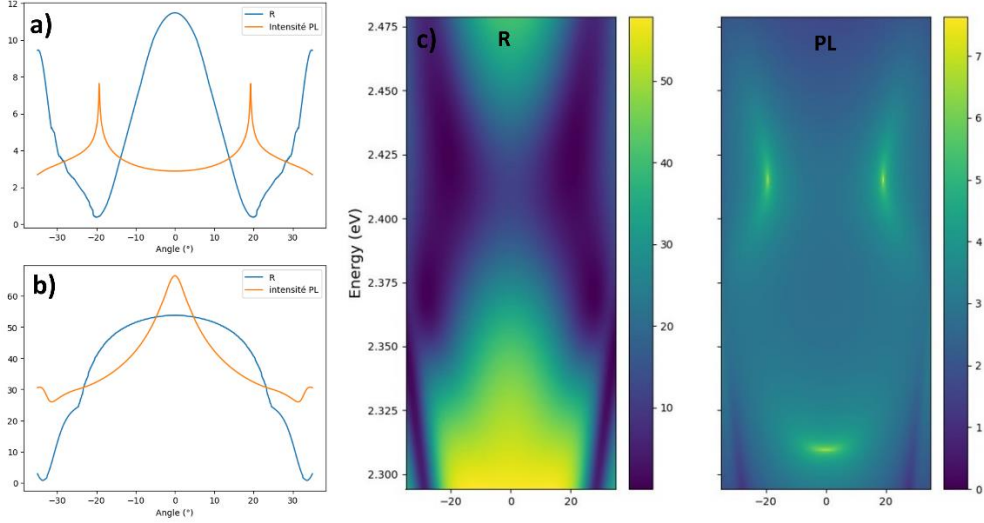

Fig. 5. (a) UP and (b) LP emission from pol-BICs. (c) Angle-resolved Pol-BIC reflectance and emission for $d_{QDs} = 94$ nm, with $\lambda_{DBR} = 450\ nm$ and fixed $d_{defect} = 590\ nm$.

Subsequently, as previously discussed, we considered the critical coupling strength, where the quantum exceptional point of this coupled system results in Rabi splitting. As previously mentioned, in typical systems, that is, with no BICs, this condition is given by $g_{QEP} = |\gamma_{cav} - \gamma_{ex}|/4$, considering zero $k$-detuning. However, for a qBIC, and then the consequent formation of a pol-BIC, $\gamma_{cav}, \gamma_{ex} \to 0$, and the previous condition for strong coupling is now given by $(\omega_{ex} - \omega_{cav})/2 = g_{QEP}$. On the other hand, the relation dispersion for the cavity mode may be described in terms of the angular position (wavevector $k$) as $\omega_{cav}(\theta) = \omega_{cav}(0)\left(1 - \frac{sin^2(\theta)}{n_{eff}^2}\right)^{-\frac{1}{2}}$, where $\omega_{cav}(0) = \omega_0$ is the cavity mode energy at normal incidence and $n_{eff}$ is the cavity effective refraction index. Subsequently, by combining these expressions, it follows that $\theta = \arcsin\left(n_{eff}\sqrt{1 - \frac{\omega_0^2}{(\omega_{ex}-2g)^2}}\right)$.

Evidently, $n_{eff}$, $\omega_0$, and $g$ depend on all parameters defining the interacting light-matter system: $\lambda_{DBR}$, $d_{defect}$ and $d_{QDs}$. By fixing $d_{defect}$, if $d_{QDs}$ varies not much, and since the cavity system is defined to concentrate the electric field from the incident light into the defect layer and for the case when $d_{defect} + Nd_B \gg d_{QDs}$, then it can be assumed that, in a general way, $n_{eff}(\lambda) \approx n_B(\lambda)$, the refractive index of the defect, which can be approximated in terms of a Cauchy relation as $n_B(\lambda) = A + \frac{B}{\lambda^2}$. From previous simulations, it was determined that $\omega_0(\lambda_{DBR}, d_{defect}, d_{QDs}) \approx \omega_0(\lambda_{DBR}, d_{defect})$, then $\omega_0(\lambda_{DBR})$, and that the variation is quite small around the bandgap of the perovskite system considered in this study (2.4 eV). For the coupling constant, $g$, a linear dependence on $d_{QDs}$, for thicknesses in the range of 90-110 nm, can be found from our simulations: $g(d_{QDs}) = Cd_{QDs} + D$. Therefore, from all these arguments, the relation dispersion for the pol-BICs, for $d_{defect}$ fixed, as a function of the parameters defining the coupled system, can be established as

$$\theta(\lambda_{DBR}, d_{QDs}) \approx \arcsin\left((A + \frac{B}{\lambda_{DBR}^2})\sqrt{1 - \frac{\omega_0^2}{(\omega_{ex}-2(Cd_{QDs}+D))^2}}\right), \quad (5)$$

where $A \approx 1.19155$, $B \approx 14\,000$, $C \approx 1.4 \times 10^{-3}$, $D \approx -0.0785$, for the perovskite system considered in this work and $d_{defect} = 590\ nm$.

Thus far, by varying the parameters of the systems responsible for the strong coupling, using values just slightly apart from the usual ones, the presence of qBICs modes for the cavity reflectance was evident. These qBICs allowed us to obtain pol-BICs with intensities greater than $10^5$. To understand and control the formation of these pol-BICs, parameter variation allows us to obtain an empirical relationship, as given by Eq. (5). To explain and formalize this empirical expression and then comprehend these pol-BICs, let us consider a different point of view, that is, the EPs expression due to the presence of qBICs.

From the strong coupling model given above, an EP occurs when the square-root term in Eq. (3) is zero. However, when a qBIC is present and a pol-BIC is formed, then $\gamma_{cav}, \gamma_{ex} \to 0$, and the emission intensity is divergent, which can be considered as a population inversion condition, for a system with a non-zero cavity-detuning. Therefore, the condition becomes $g^2 + \delta^2 = g^2 + (\omega_{cav}(\theta) - \omega_{ex})^2 = 0$. The EPs that give rise to the coupling condition are those identified as EP2 in Ref. [22] and the corresponding coupling constant becomes imaginary:

$$g = i\delta. \qquad (6)$$

From this equation, it can be swiftly shown that all the pol-BICs obtained for $d_{defect} = 590\ nm$ and $\lambda_{DBR} = 480\ nm$, and a variation in $d_{QDs}$ from 92 nm to 112 nm, which can be correctly fitted using parametric and empirical Eq. (5), can be exactly predicted by the strong coupling model, provided that the conditions for Eq. (6) were validated. This is shown in Fig. 6 for some of those cases, finding a dispersion relation able to correctly fit the mixed polaritons obtained from these pol-BICs, and to show the hybridization of the pol-BICs formed in these interactions. In the supplementary information, GIFs images show the polariton formation for real and imaginary $g$, for both possible cases of non-zero cavity-detuning. Therefore, in contrast to real $g$, a hybridization scheme for imaginary $g$ is proposed, as shown in Fig. 7, for positive cavity-detuning, which is in direct agreement with the GIFs images for the corresponding cases.

This hybridization scheme can be deduced by analyzing the response of a two-level atom to an optical field that is sufficiently intense to remove a significant fraction of the electron population from the atomic ground state. From the corresponding Schrödinger equation, adapted accordingly to our case, that is, to consider the perovskite two-level QDs and their interaction with the cavity mode, and using the rotating-wave approximation, a specific expression for the wavefunction can be found, dependent on the initial conditions [53]. By considering as the initial condition that the probability of being in either the ground or excited state of the QDs is constant in time, one can obtain time-independent probabilities of occupancy for these states, leading to dressed states, which are orthogonal, stationary solutions of the initial Schrödinger equation, despite not being energy eigenstates. To obtain these states, a complex Rabi frequency is typically considered in textbooks [53]. The usual hybridizing scheme for real $g$ and that shown in Fig. 7 for imaginary $g$ represent the complete set of dressed states, that is, the stationary states of the coupled emitter-field system [53,54]. Therefore, for all these states to manifest in strong coupling, Eq. (6) must be considered. An additional advantage of considering the full extent of strong coupling, by admitting real and imaginary coupling constants, is the possibility of observing a Mollow triplet [55], as is hinted in Fig. 6 when considering $d_{QDs} = 112\ nm$: several emission peaks can be observed at an angle of approximately 28°. This feature could be very relevant for applications in quantum technology.

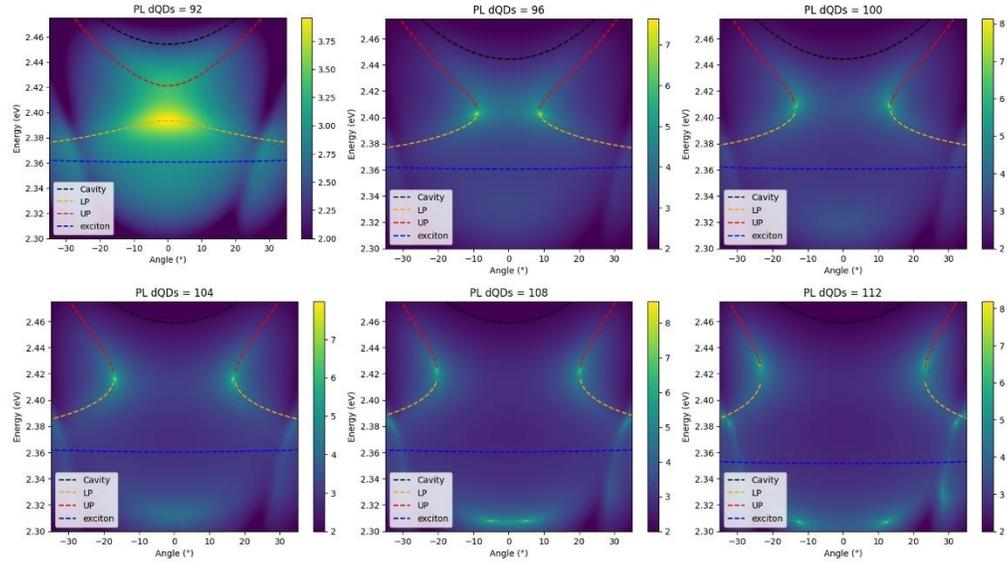

Fig. 6. Relation dispersion in logarithmic scale of mixed polaritons for several thicknesses of the perovskite layer, $d_{QDs}$, with $\lambda_{DBR} = 480\ nm$ and fixed $d_{defect} = 590\ nm$.

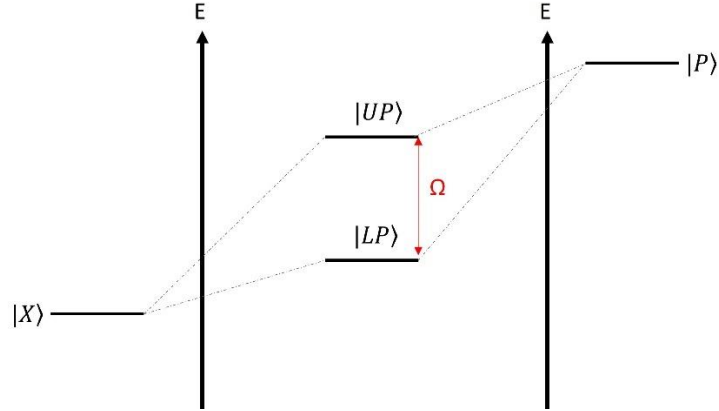

Fig. 7. Hybridization scheme for mixed polaritons, case of positive cavity-detuning.

However, not all pol-BICs are mixed-polaritons. Several situations for different types of pol-BICs can be distinguished from our previous simulations, specifically from our previous cartography (Fig. 4), by examining their corresponding dispersion relations. Fig. 8 illustrates these possible situations: i) normal polaritons, which are usually described as LP and UP (yellow spots); ii) mixed polaritons, which can be fitted exclusively using Eq. (6), as illustrated in Fig. 6 (red spots); and iii) other polaritons, which can be found in two cases: the confluence of different modes creating polaritons with several shapes, or polaritons with both high intensity and large spectral width (black spots). The final case is illustrated in Fig. 9 on a logarithmic scale.

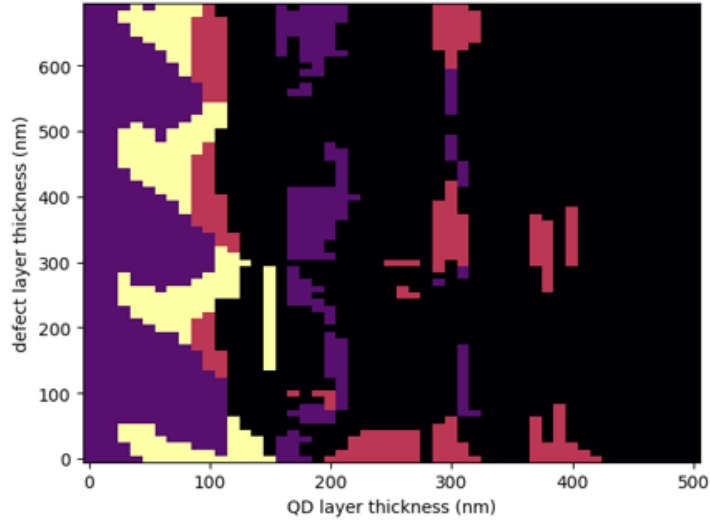

Fig. 8. Cartography of different types of pol-BICs: yellow, normal LP and UP polaritons; red, mixed polaritons; black, other polaritons like those surging from the confluence of several modes or those with high intensity and large spectral width Violet spots correspond to uniform intensities, that is where no coupling is observed.

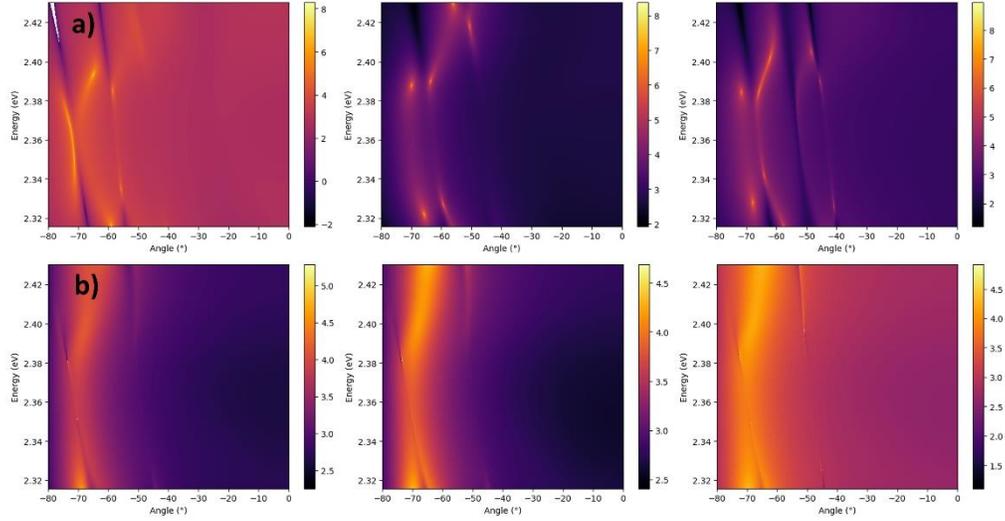

Fig. 9. Other polaritons like (a) those surging from the confluence of several modes, or (b) those showing high intensity and large spectral width.

## 3.3 Topological aspects of mixed polaritons

As indicated in [31], pol-BICs have a topological nature inherited from qBICs, which are at the origin of their formation, even though these qBICs have a finite quality factor. For 1D photonic systems, the topological nature of the mixed polaritons studied in this work is characterized by the Zak phase [56], which can be calculated according to the surface bulk correspondence as $\exp(i\theta_n^{Zak}) = -\frac{\text{sgn}(\phi_n)}{\text{sgn}(\phi_{n-1})}$, which can have values of 0 or $\pi$, and $\phi_n$, $\phi_{n-1}$ are the reflection phases in the bandgaps above and below the $n_{th}$ band [38]. To calculate the Zak phase for our system, first, because the cavity is not a symmetrical structure, the band structure was calculated by considering all the sequence as the unit cell. Because the actual system is composed of only one unit cell, a layer of air among the unit cells was also considered in the calculation, with a thickness of 1 μm. From the reflectance and reflection phases calculated under this assumption, as shown in Fig. 10(a) for qBIC, it is evident that there is a discontinuity in the Zak phase, which introduces an interface state in the gap between the bands where this discontinuity occurs, also indicating band inversion, which is a topological transition [38,56]. The gap where the discontinuity is observed, that is, the change from 0 to π, is centered at 2.4 eV, that is, the gap tuned to the exciton of the perovskite QDs. The discontinuity in the Zak phase has been recognized as a signature of the existence of a topological mode, which is evidenced by the rigorous relationship between the Zak phase and reflection phase given by the previous equation and its corresponding numerical or experimental verification [56]. The analysis was preserved when the coupled system was considered, as shown in Fig. 10(b). It is worth noting that the discontinuity is not perfect because we are dealing with imperfect BICs, that is, qBICs. Nevertheless, this topological transition, given by band inversion, marks the appearance of exceptional points when the Zak phase is zero in its transition from the positive to negative values of the reflection phase, that is, at the interface state in the gap. Therefore, the topological nature of mixed polaritons reinforces the modeling of their dispersion relation using Eq. (6).

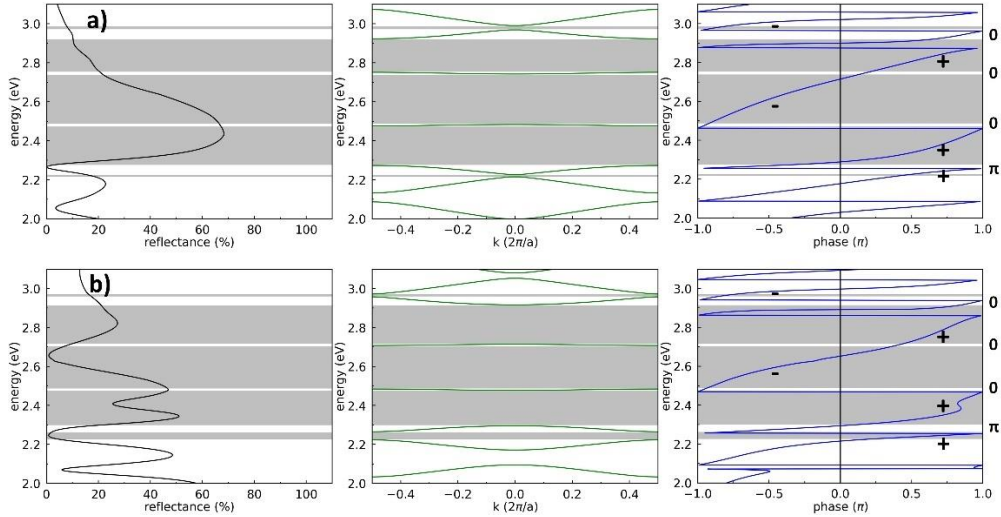

Fig. 10. From left to right, reflectance, band structure and reflection phase for (a) the asymmetric cavity and (b) the coupled system. Zak phase discontinuity is illustrated in the reflection phase for both cases.

*3.4 Experimental results*

From the polariton classification shown in Fig. 8, it can be observed that obtaining experimental evidence for mixed polaritons requires precise sample fabrication. Otherwise, normal polaritons are obtained in the best case. Slight deviations from a given set of fabrication parameters, $(\lambda_{DBR}, d_{defect}, d_{QDs})$, can result in mixing of different types of polaritons, making it difficult to correctly classify them. Therefore, experimental evidence of mixed polaritons is still in progress, and the experimental results presented below show their first glimpse.

A sample with parameters $(\lambda_{DBR} = 480\ nm, d_{defect} = 590\ nm, d_{QDs} = 50\ nm)$ was fabricated. First, electron microscopy experimental conditions were found to provoke the evaporation of some parts of the perovskite QDs thin film deposited on the p-Si cavity surface. From the resulting islands, a thickness of 50 nm was measured and used in the corresponding simulations (see corresponding images in the supplementary information).

Because our ARPL setup is based on emission acquisition using an optical fiber, instead of the typical Fourier-Transform one using a microscope objective, the numerical aperture of the fiber naturally increases the wavevector or angular data dispersion, explaining the diffuse shape of the normalized emission shown on the right side of Fig. 11(b). It is also worth noting that the ARPL measurements clearly show a shift of more than 30 nm in the emission peak, providing evidence of the polariton formation, as no shift was observed for the perovskite QDs solution when varying the angular position of the optical fiber. Consequently, the dispersion relations to fit the simulated and experimental normalized data must consider a non-zero coupling, either real or imaginary. As observed in Fig. 11(c) and (d), a set of three pairs of polaritons can be used to fit the simulated and experimental data, with one pair always being a normal polariton, and a real coupling corresponding to a Rabi splitting of 28 meV. For the other two pairs, Fig. 11(c) shows the fit obtained using normal polaritons, with a real coupling corresponding to a Rabi splitting of 16 meV for both; while Fig. 11(d) shows the fit obtained using mixed polaritons, with an imaginary coupling corresponding to a negative Rabi splitting of -16 meV. From these figures, (c) and (d), to our best judgement, mixed polaritons allow, in general, the best fitting of the simulated or measured emission, but, in particular, of the region marked by the arrow. As mentioned before, this work is still in progress, and more strong evidence of mixed polaritons is required.

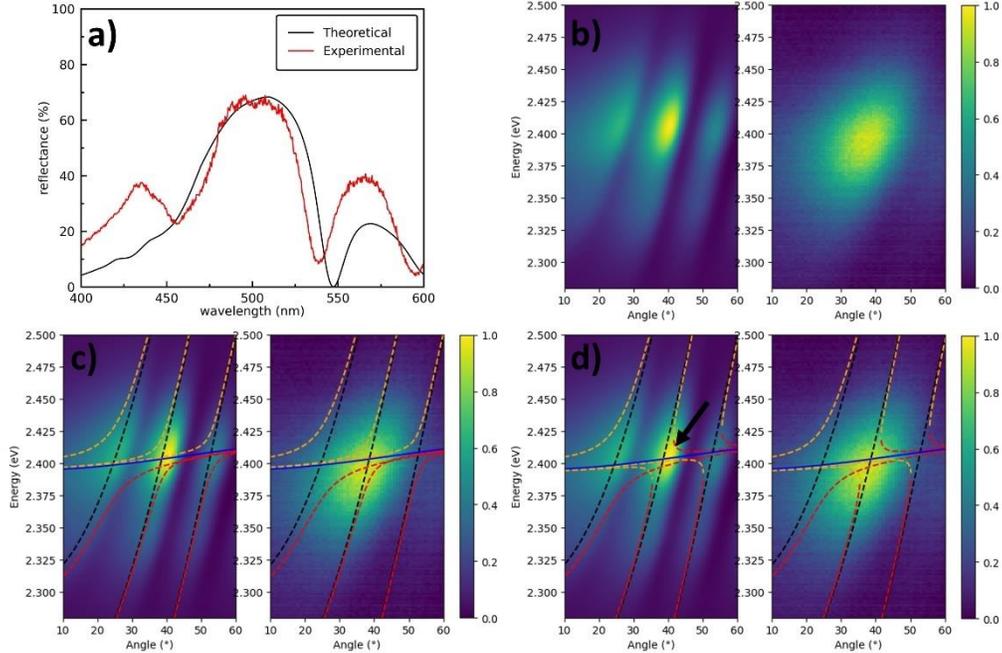

Fig. 11. Experimental results for M1 sample. a) Reflectance at normal incidence. b) ARPL best-fit simulation for $d_{QDs} = 32$ nm (left) and measurement (right) for sample M1. Dispersion relationships using c) normal (real $g$) and d) mixed (imaginary $g$) polaritons. Blue dashed line represents the exciton dispersion, black dashed line the main cavity mode, and orange (red) dashed lines represent the upper (lower) polaritons.

## 4. Conclusions

In this study, using the physical properties of quasi–Bound States in the Continuum to provoke the manifestation of correct Exceptional Points, we present numerical and experimental evidence of mixed polaritons, for which an imaginary coupling coefficient is necessary. From the numerical approach, an empirical parameterizing relationship can be deduced, which is related to the strong coupling model. Using its non-Hermitian quality, the complex character of the corresponding coupling coefficient for mixed polaritons was evident.

To support the appearance of the correct EPs giving place to the imaginary coupling coefficient, a topological transition was made evident using the Zak phase. This topological transition, caused by band inversion, occurs when the reflection phase changes from zero to π. This discontinuity introduces an interface state in the gap between the bands where it occurs, which in this case corresponds to that directly related to the exciton used to form polariton quasiparticles. Therefore, the topological nature of mixed polaritons reinforces the modeling of their dispersion relation using Eq. (6).

Finally, experimental evidence of these mixed polaritons was obtained at room-temperature for a $CsPbBr_3$ perovskite system using p-Si microcavities. The simplicity of the porous silicon photonic structure used to obtain these results is worth noting. The structure was fabricated without any special characteristic conditioning of reproducible strong coupling of the photonic cavity with this emissive system at room temperature. Despite its natural disorder and many structural defects on the nanometer scale, its topological features make it robust and stable for strong coupling with quantum emitters in a systematic manner. The main parameter to be carefully controlled during the fabrication process being the thickness of the perovskite QDs layer. Thus, there is room for improvement in the porosity of these photonic structures, their interactions with excitonic systems, and the resulting strong coupling between them.

**Funding.** ECOS-Nord CONACyT-ANUIES 315658; National Science Foundation (456789); PAPIIT-UNAM IN112022. PASPA-UNAM.

**Acknowledgments.** This research was partially funded by the ECOS-Nord CONAHCyT-ANUIES 315658, PAPIIT-UNAM IN112022, and PAPIIT-UNAM IN109122. J.A.R.E. is grateful for the sabbatical funding from PASPA-DGAPA-UNAM, CONAHCyT, and the University of Sherbrooke. S.E.G. thanks CONAHCyT for postdoctoral fellowship. V. F. thanks MITACS and X scholarships. The authors wish to acknowledge the technical assistance provided by Gerardo Daniel Rayo López and Gabriel Laliberté; Rogelio Morán Elvira, José Campos Álvarez and Carlos Magaña Zavala with the SEM images. LN2 is an International Research Laboratory (IRL) funded and co-operated by Université de Sherbrooke (UdS), Centre National de la Recherche Scientifique (CNRS), Ecole Centrale Lyon (ECL), Institut National des Sciences Appliquées de Lyon (INSA Lyon), and Université Grenoble Alpes (UGA). The work was also financially supported by the Fond de Recherche du Québec Nature et Technologies FRQNT and CMC Microsystems.

**Data availability.** The data underlying the results presented in this study are available upon reasonable request.

**Supplemental document.** See Supplement 1 for supporting content.

## 5. Supplementary Information

Polariton formation for real and imaginary $g$. Positive and negative detuning.

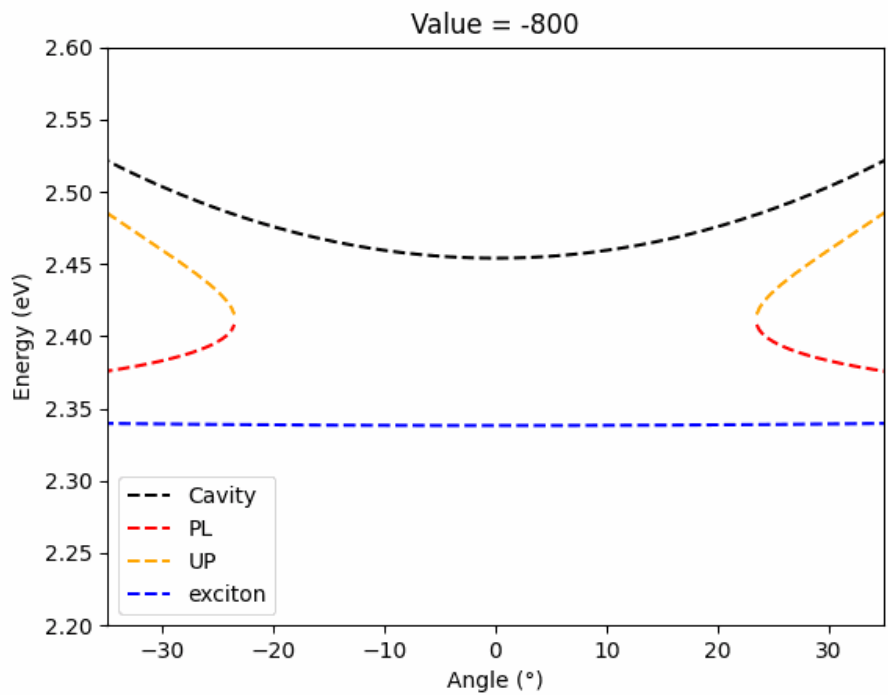

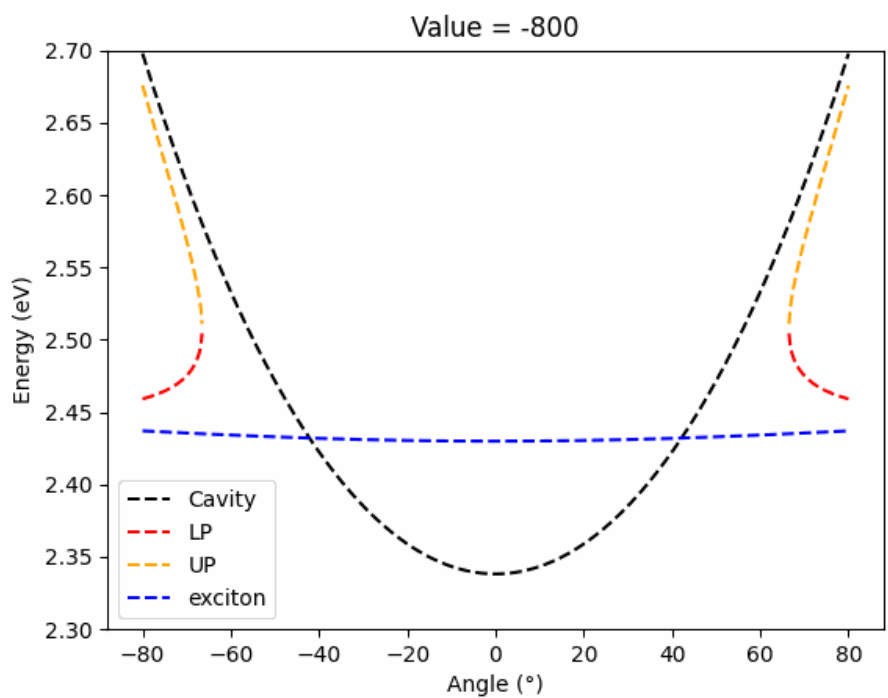

Fig. S1. Real and imaginary $g$. Positive and negative cavity-detuning.

Electron micrographs taken by using SEM .

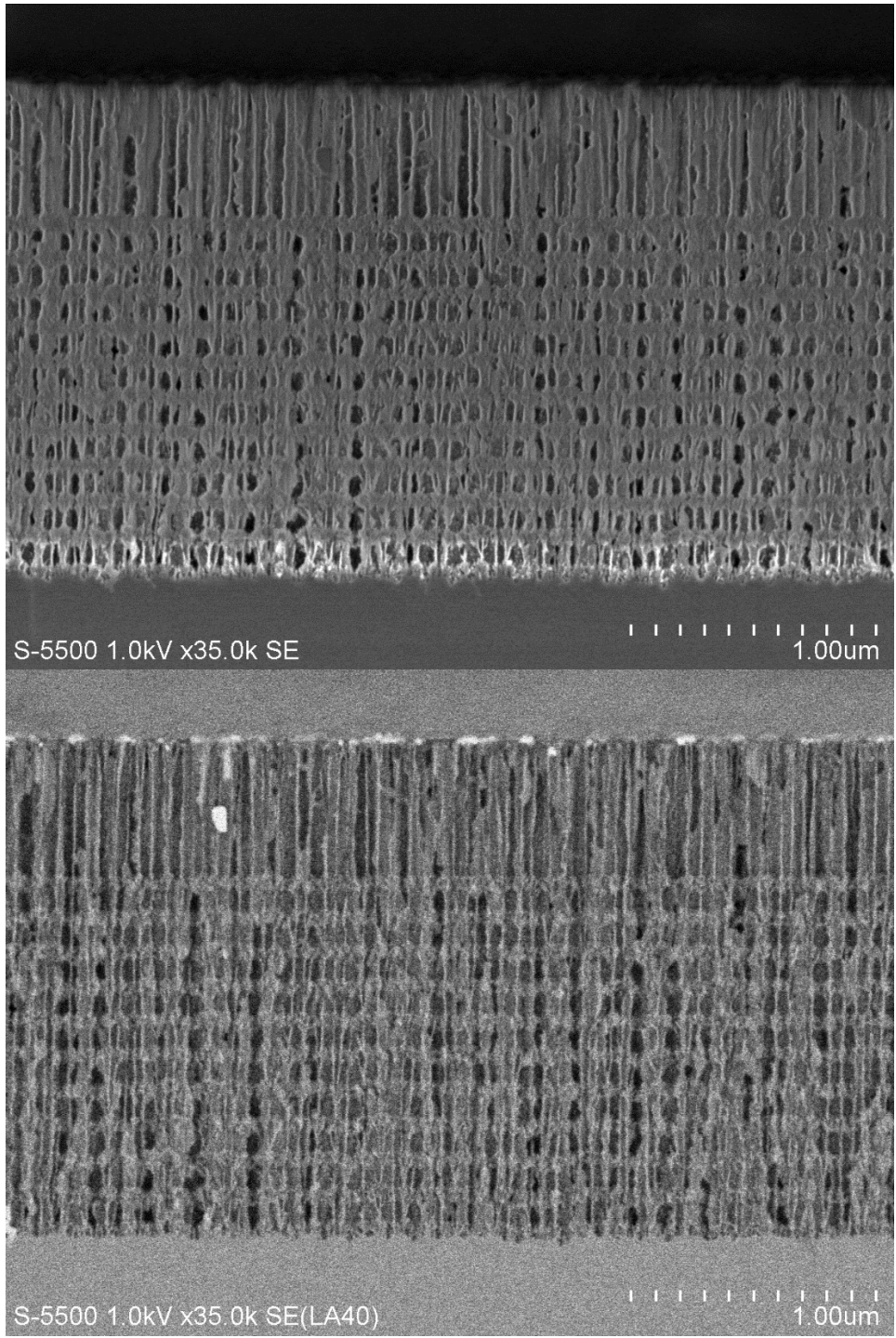

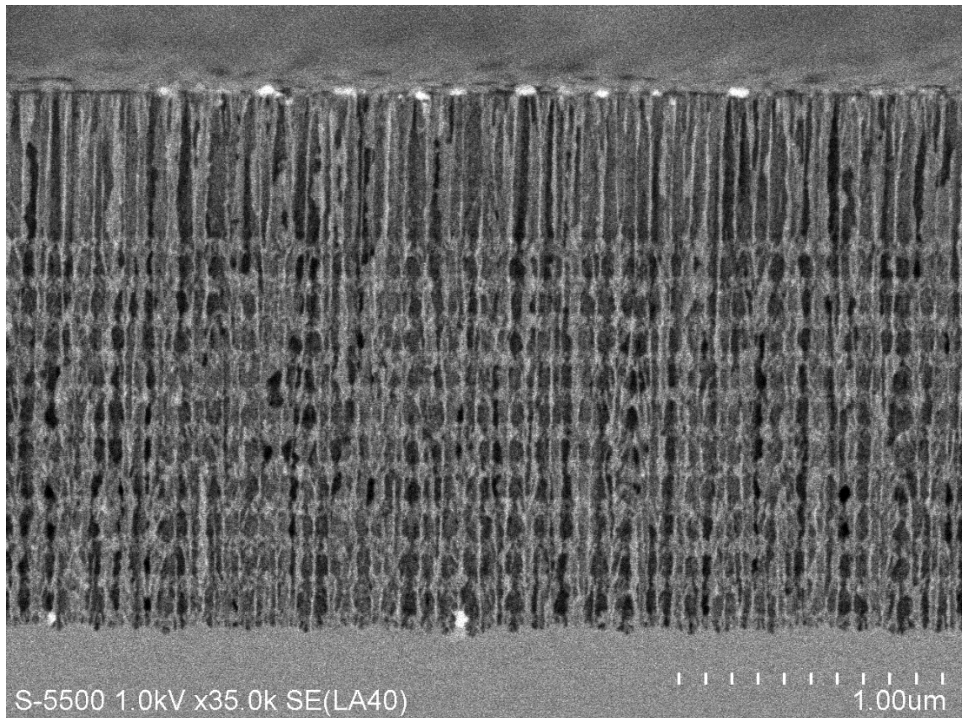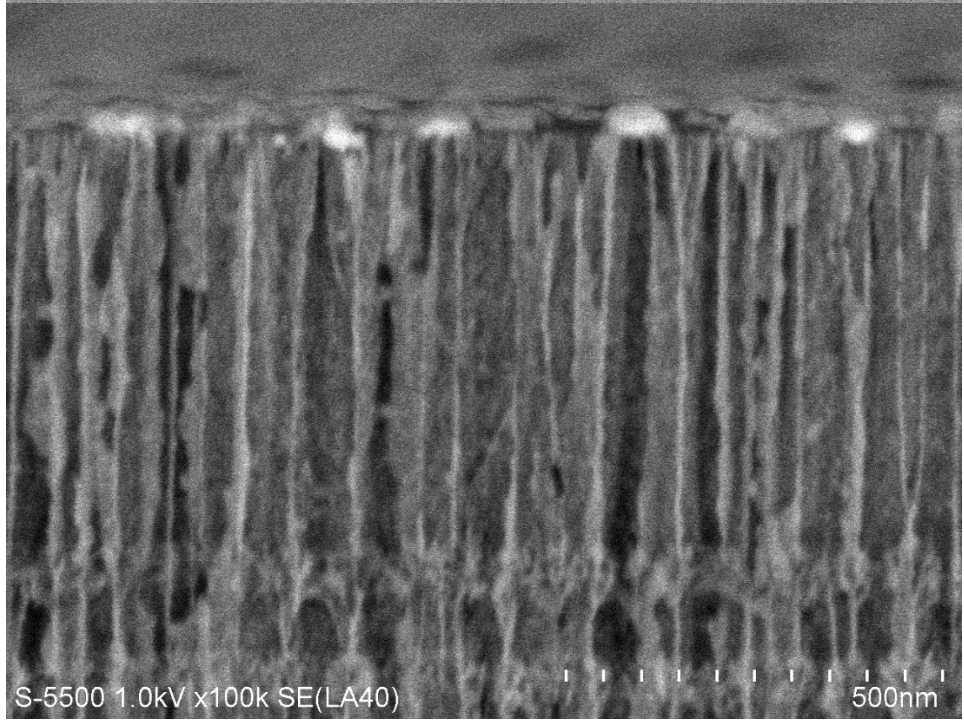

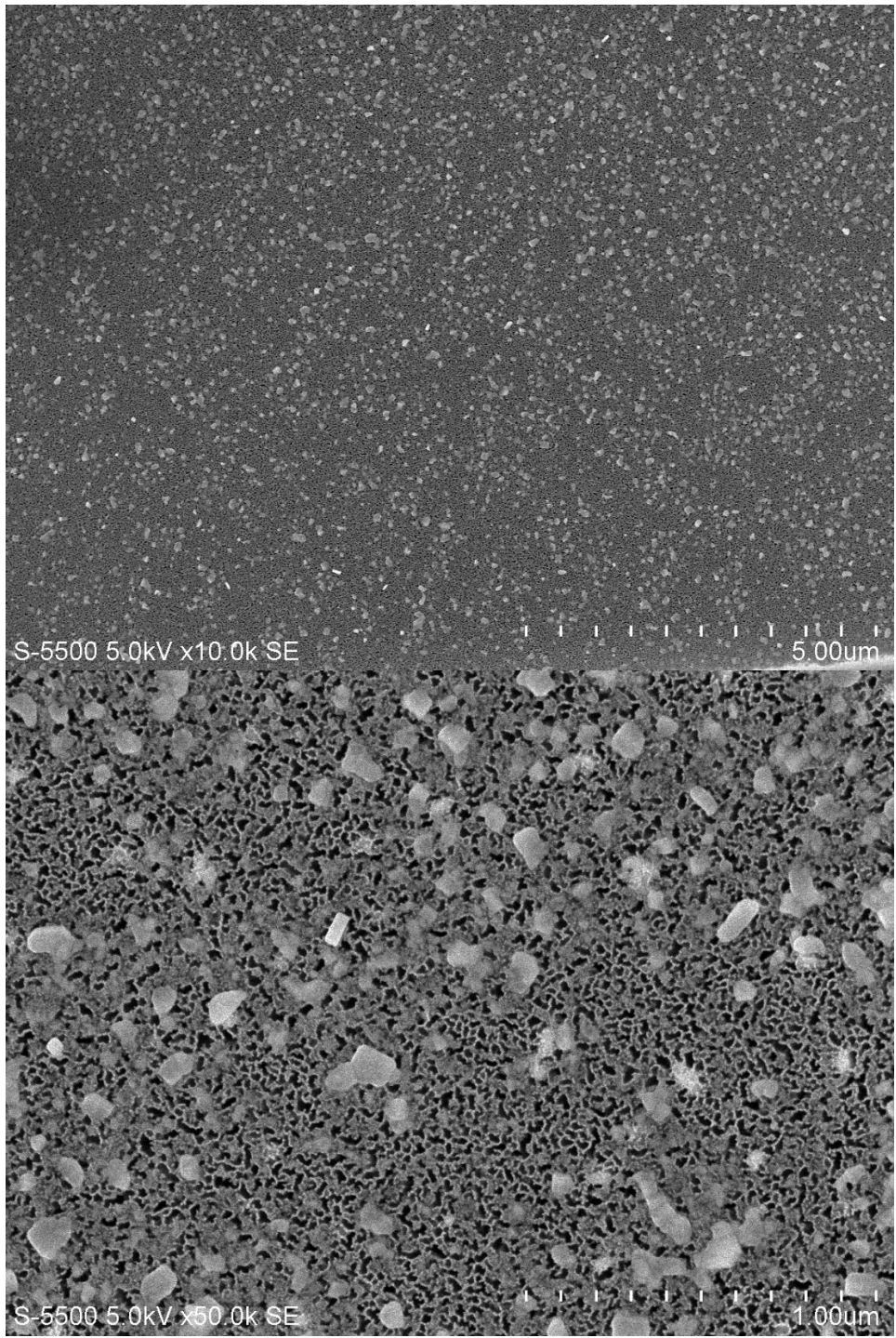

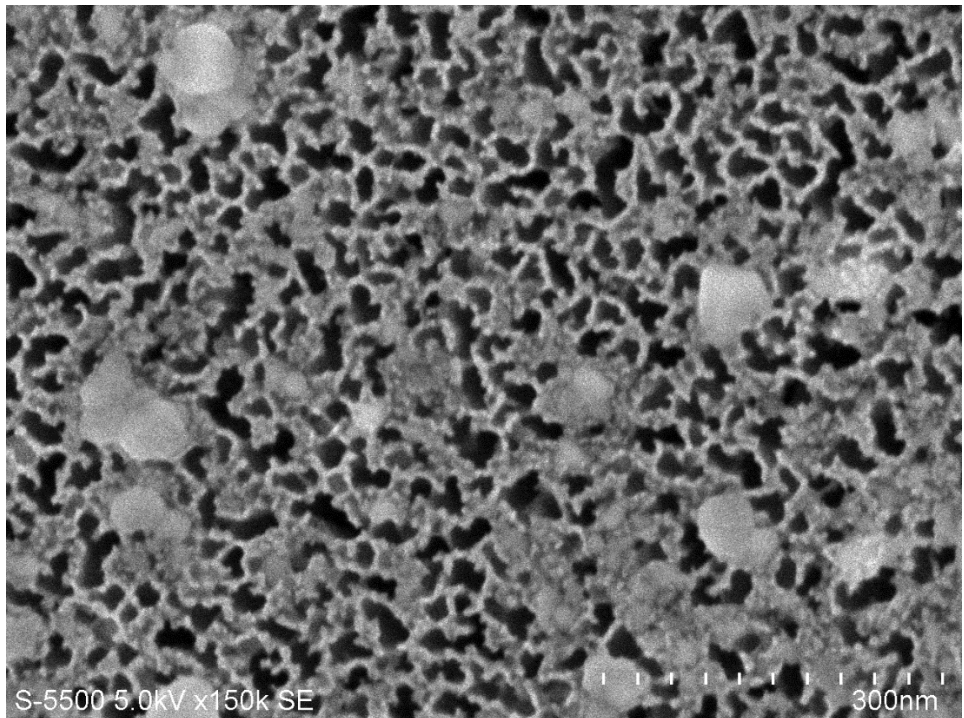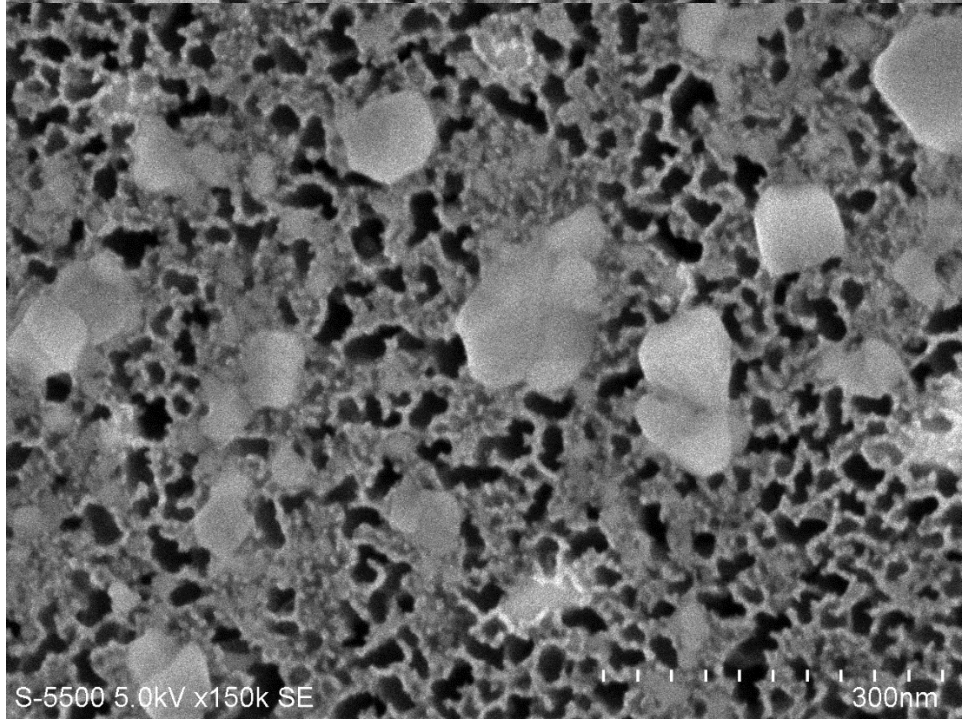

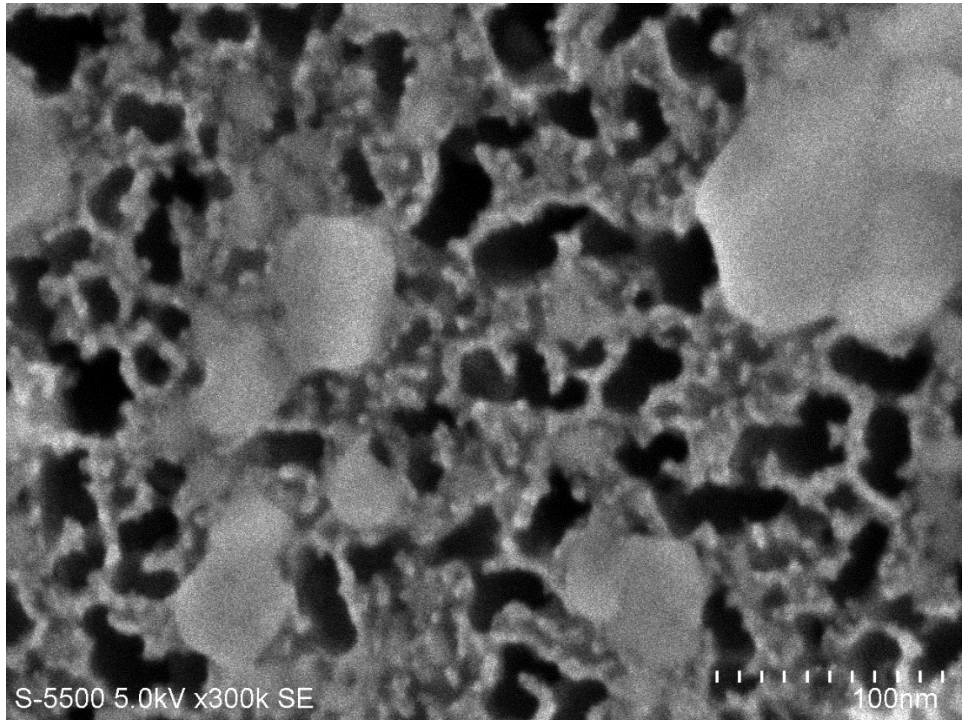

Fig. S2. SEM micrographs from the cross section and surface of the coupled system, showing the perovskite islands formed from the deposited film because of the action of the electron beam.